# R-process Experimental Campaign at the National Superconducting Cyclotron Laboratory

**Jorge Pereira[1,2], Ana Becerril[1,2], Thom Elliot[1,2], Alfredo Estrade[1,2,3], Daniel Galaviz[1,2], Linda Kern[1,2], G. Lorusso[1,2,3], Paul Mantica[1,4], Milan Matos[1,2], Fernando Montes[1,2,3], Hendrik Schatz[1,2,3]**

*1) National Superconducting Cyclotron Laboratory (NSCL),*
*2) Joint Institute of Nuclear Astrophysics (JINA),*
*3) Department of Physics and Astronomy,*
*4) Department of Chemistry*
*Michigan State University (East Lansing) Michigan, USA*
*E-mail:* *pereira@nscl.msu.edu*

**Stefan Hennrich[1,2], Karl-Ludwig Kratz[1,2], Oliver Arndt[1,2], Ruben Kessler[1,2], Florian Schertz[1,2], Bernd Pfeiffer[1,2]**

*1) Institut für Kernchemie,*
*2) Virtuelles Institut für Struktur der Kerne and Nuklearer Astrophysik (VISTARS)*
*Johannes Gutenberg, Universität Mainz (Mainz) Germany*

**Matt Quinn[1,2], Ani Aprahamian[1,2], Andreas Woehr[1,2]**

*1) Institute of Structure and Nuclear Astrophysics,*
*2) Joint Institute of Nuclear Astrophysics (JINA)*
*University of Notre Dame (South Bend) Indiana, USA*

**Ed Smith[1,2]**

*1) Department of Physics*
*Ohio State University (Columbus) Ohio, USA,*
*2) Joint Institute of Nuclear Astrophysics (JINA)*
*Michigan State University (East Lansing) Michigan, USA*

**William Walters**

*Department of Chemistry and Biochemistry*
*University of Maryland (College Park) Maryland, USA*

A JINA/VISTARS r-process campaign was completed at the A1900 Fragment Separator of the National Superconducting Cyclotron Laboratory in the fall of 2005. The purpose of the campaign was the measurement of the β-decay half-lives and β-delayed neutron-emission probabilities of different unknown neutron-rich nuclei participating in the r-process. Details of this campaign are presented.

**1. Introduction**

A comprehensive theoretical and experimental knowledge of the structure of nuclei far away from stability is mandatory in order to fix the astrophysical properties of the r-process scenario(s) [1]. An illustrative example of the dichotomy Astrophysics vs. Nuclear Physics is found in the region prior to the A=130 abundance peak, where r-process models tend to underestimate the productions by an order of magnitude or more. On one hand, it is claimed that the lower part of the A=130 abundance peak is sensitive to neutrino post-processing effects [2]. On the other hand, it has been demonstrated that this region is very sensitive to the details of the structure of the corresponding r-process nuclei; in particular, it has been shown that the underproduction of abundances can be largely corrected if one assumes a reduction or *quenching* of the N=82 shell gap far from stability [3-5]. Therefore, disparities between r-process model calculations with observed abundances provide unique insight about the neutrino flux during the r-process, once the underlying nuclear structure is understood.

**2. Experiments**

A JINA/VISTARS r-process campaign was completed at the A1900 Fragment Separator of the National Superconducting Cyclotron Laboratory in the fall of 2005. A first experiment, was focused on the N≥54 Ge to Br isotopes, which can constitute part of the N>50 neutron-rich seed nuclear composition after α-rich freeze-out, in the hot-bubble neutrino-wind scenario [6]. From a nuclear-structure point of view, these nuclei lie in the region between the N=56 sub-shell closure and the *sudden onset of deformation* at N=60, south of the range Sr-Zr, for which the most pronounced transition from spherical to strongly deformed ground-state shapes has been observed. A second experiment was focused on the nuclear structure of A≈110 r-process nuclei. Besides the shape evolution from the sub-shell N=56 to the *sudden onset of deformation* at N=60, the region of the refractory elements around Zr beyond N≈66 should contain another phase transition from strongly prolate, oblate or triaxial shapes to spherical neutron magicity at N=82. Self-consistent HFB calculations of the potential-energy surface (PES) of neutron-rich Zr isotopes [7,8], however, have shown that the tetrahedral minimum in the PES of $^{110}$Zr and $^{112}$Zr may lie lower than their secondary prolate minimum. The experimental observation of this local spherical "tetrahedral magic gap" near N=70 would have strong consequences in the understanding of the r-process, as it may be an indirect signature of the N=82 *shell quenching*.

The experimental observables measured in the completed experiments were the β-decay half-lives and β-delayed neutron-emission probabilities. These particular observables were chosen because 1) they are direct inputs in r-process model calculations which will substitute theoretical estimates, and 2) as both experiments deal with the degree of deformation of neutron-rich nuclei in the r-process, a first hint about the shape of a nucleus can already come from the measurement of the β-decay half-life ($T_{1/2}$) and the β-delayed neutron-emission probability ($P_n$).

The campaign was performed at the National Superconducting Cyclotron Laboratory (NSCL) at Michigan State University (MSU). The two coupled cyclotrons (K500 + K1200), at the beginning of the beam line, accelerated a $^{136}$Xe primary beam up to 120MeV/u. This beam was then transmitted into a 242mg/cm$^2$ Be target located at the entrance of the A1900 in-flight separator. The nuclei investigated in each of the two experiments were produced by fragmentation of the $^{136}$Xe beam by beryllium nuclei. Due to momentum conservation, the reaction products were forward-emitted into the A1900. An achromatic Al wedge, mounted in the dispersive focal plane, combined with the optics of the A1900, allowed a spatial separation of the nuclei transmitted to the final focal plane according to their different energy losses and magnetic rigidities. Two position-sensitive plastic scintillators at the intermediate (dispersive) and final (achromatic) focal planes were used to measure the time-of-flight (ToF) of the transmitted nuclei, as well as to correct the intrinsic dependence of this time-of-flight with their transversal position at the dispersive focal plane. Once separated, the nuclei were guided into the final experimental area, equipped with an implantation setup designed to measure β-decay half-lives and β-delayed neutron-emission probabilities of implanted nuclei. By combining the energy-loss signals, measured with a Si detector at the experimental area, with the ToF signal along the A1900, it was possible to separate unambiguously the different nuclear species coming in the same cocktail.

In the first part of the experiment (around 8 hours of beam time), known microsecond isomers were implanted in a 4 mm Al degrader surrounded by three Segmented Germanium Array (SeGA) detectors [9], symmetrically mounted around the beam axis. By measuring the gamma lines of these isomers it was possible to define the particle identification of the transmitted nuclei. The β-decay properties of these nuclei were measured in the second part of the experiment: The NSCL Beta Counting System (BCS) [10] was placed downstream of the gamma-ID station. It consisted of a stack of four Si PIN detectors, used to measure the energy-loss of the exotic species, followed by a doubly segmented 40×40 Si strip detector (DSSD), where the nuclei were implanted. A final single-segmented 16 Si strip detector (SSSD) was used to veto light particles and nuclei that punch-through the implantation detector. The BCS was surrounded by the Neutron Emission Ratio Observer (NERO) detector [11], which was used to measure β-delayed neutrons. This detector consisted of 16 $^3$He and 44 B$_3$F proportional counter tubes embedded in a polyethylene matrix used to thermalize the emitted neutrons.

## 3. Analysis

In order to separate implantation, decay and background events, specific conditions were assigned to each of them. When a signal was registered in the first Si detector and in the DSSD, the event was considered to be an implantation. A decay event was assumed when there is a coincidence between no-signal in the first Si detector and signals from the DSSD. When an implantation event is measured, the time –recorded from a 50MHz VME SIS3820 clock– and the pixel at which the event was implanted were recorded. If a consecutive decay event occurs in the same pixel within a time gate of 10 seconds, then the event is associated with the previous implantation event and its time is registered too. By distributing the time differences between implantation and decay events associated with a given nucleus, we could produce a decay-curve. Figure 1 shows the preliminary decay-curve obtained for $^{105}$Zr and $^{106}$Zr. Multi-parameter fits including background and decay-event curves from mother, daughter and granddaughter nuclei are currently being calculated to extract half-lives.

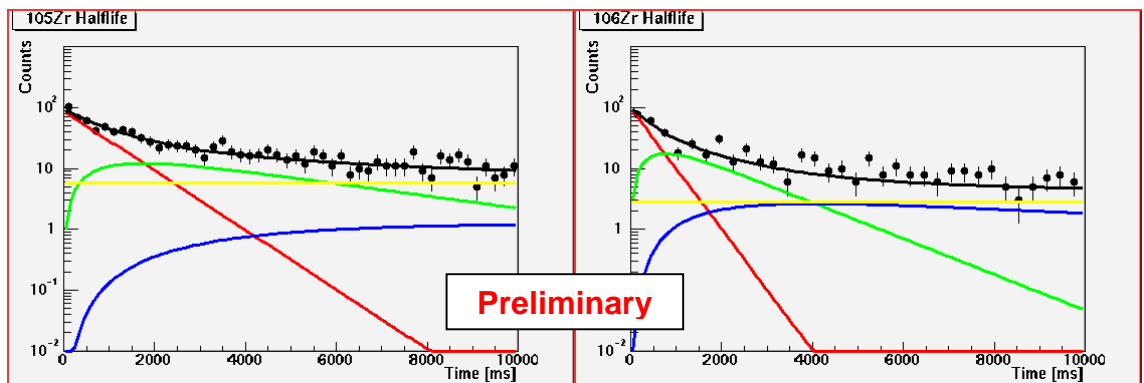

**Fig.1:** *β-decay curve of $^{105}$Zr and $^{106}$Zr measured during one of the experiments. The lines correspond to fits done in order to disentangle the contributions from the decaying parent nucleus (red), daughter (green), granddaughter (blue) and background (yellow). The black line corresponds to the sum of the different contributions.*

After a decay event was identified, a gate of 200 μsec was opened in order to search for possible correlated delayed neutrons. This time window was calculated on the basis of the time needed by the emitted neutrons to be thermalized in the polyethylene matrix before being detected in one of the proportional counters. By grouping the neutron events according to their multiplicities and normalizing them to the total number of β-decay associated with the corresponding decaying nucleus, we can then extract the $P_n$-value.

The data analysis is currently being done in collaboration between the National Superconducting Cyclotron Laboratory (Michigan), the University of Notre Dame (Indiana) and the University of Mainz (Germany).

**4. Summary**

An r-process motivated experimental campaign was carried out at the National Superconducting Cyclotron Laboratory. The goal of the campaign was to investigate the shape-evolution of nuclei far away from stability that are involved in the r-process. The probes used for such studies were the β-decay half-lives and $P_n$-values, which are particularly sensitive to the structure of the decaying nuclei at different energies through the β-strength function. Results obtained from the data analysis will provide insight into relevant questions about nuclear structure of exotic nuclei including *shell quenching* and pn-interactions between exotic combinations of valence nucleons. The knowledge gained from this analysis in terms of nuclear structure will put r-process models on a more solid basis, in order to investigate the astrophysical conditions that might affect the synthesis of heavy elements. The measured β-decay properties will also provide direct inputs needed in r-process model calculations.

The performance of the detection setup, including a special blocking system designed to remove primary-beam charge-state contaminants from the cocktail beam, demonstrated the capabilities of the NSCL to perform β-decay studies of extremely neutron-rich r-process nuclei. Results from previous β-decay experiments have been already published (see e.g. Refs. [12-16]). Moreover, new techniques to measure masses of neutron-rich nuclei have being recently tested [17], making the NSCL one of the prime facilities to perform this type of r-process experiments. Future next generation facilities (e.g. the Isotope Science Facility ISF at Michigan State University) will be necessary to cover unknown r-process regions, in order to solve at last and at once one of the most intriguing questions of Nature, namely, the origin of the heavy elements [18].